\documentclass{article}
\usepackage{amsmath}
\usepackage{amssymb}
\usepackage[Gray,squaren,thinqspace,thinspace]{SIunits}
\usepackage{graphicx}
\usepackage[small,bf]{caption}
\usepackage{a4wide}
\usepackage{subfig}
\usepackage{bbm}
\usepackage{feynmf}
\newcommand{\mx}{\ensuremath{\mathsf}}
\DeclareMathOperator{\tr}{tr}
\begin{document}
\title{Schwinger--Dyson equations for gauge-invariant correlators}
\author{David Vercauteren\thanks{E-mail: David.Vercauteren@UV.es}\\\\
\textit{\tiny Universitat de Val\`encia, Departament de F\'isica Te\`orica, Avda. Dr. Moliner 50, E-46100 Burjassot, Spain}}
\date{}
\maketitle

\begin{abstract}
The formalism for Schwinger--Dyson equations of Wilson loops, as developed by Eguchi, Weingarten, and Foerster, is extended to include connected correlators. These should allow one to study propagators and interactions of glueball fields without having to handle gauge dependent fields such as gluons an ghosts. Some steps towards a solution are made.
\end{abstract}

\section{Introduction}
Schwinger--Dyson equations are a powerful analytic tool to study quantum field theories beyond perturbation theory. In writing down the equations, no approximations need to be made, which means that the equations contain, in principle, all physics --- both perturbative and nonperturbative --- exactly.

In a general quantum field theory, the equations follow from Stokes' theorem applied to the path integral:
\begin{equation} \label{sdstart}
\int [d\phi] \frac\delta{\delta\phi(x)} e^{-S[\phi]+J\cdot\phi} = 0 \;.
\end{equation}
From this master equation, the entire tower of Schwinger--Dyson equations can be written down by expansion in the source field $J(x)$ and equating the expansion coefficients with zero. This leads to equations involving full (i.e. containing disconnected parts) $n$-point functions, and with some further manipulations they can be reduced to equations involving only connected or only 1PI Green's functions.

In the case of Yang--Mills theory (and, by extension, of QCD) the role of the field $\phi(x)$ is played by the gluon field $A_\mu^a(x)$. This means that the Schwinger--Dyson equations obtained in this way will contain gluon propagators and gluon interaction vertices. But what, in the end, interests us most are not so much gluons but rather bound-states like glueballs (and, in QCD, hadrons), meaning that after solving the full tower of equations --- supposing that we are smart enough to do so --- we would still be facing the formidable task of putting all elements together to extract the physical quantities of interest. One can therefore wonder whether it would not be possible to write down a set of equations involving only glueballs, thus sidestepping at least one leg in the quest to a better understanding of Yang--Mills theory.

This leads us to the question of the gauge symmetry. The gluon field has a nontrivial gauge transform, which makes it rather tough to get rid of the gauge dependence when starting from something like \eqref{sdstart}. Ideally, one would like to take the functional derivative in some gauge invariant way, such that only gauge invariant quantities show up in the resulting equations. These gauge invariant operators would then be interpreted as glueball operators, and the goal mentioned above would be reached.

It seems, however, that such an ambitious program does not come easily when working in the continuum. Work in that direction has been done in times past \cite{Nambu:1978bd,Corrigan:1978zg,Gervais:1978mp,Makeenko:1980vm}, but it seems like results are come by most easily when introducing a lattice regulator. This means that manifest Lorentz invariance will have to be sacrificed, to be subsequently recovered when we are able to take the continuum limit. All work done in that direction in the past (see, for example, \cite{Eguchi:1979nk,Weingarten:1979zj,Forster:1979wf,Friedan:1980tu}) focused on computing expectation values of Wilson loops, hoping to get a better understanding of confinement. For a review see \cite{Migdal:1984gj} and references therein. Section \ref{sdotl} reviews the formalism of Schwinger--Dyson equations for Wilson loops on the lattice.

In my work I aim for connected correlators, which should give more insight in the spectrum of the theory and in scattering amplitudes. In section \ref{cte} the equations for different connected and 1PI $n$-point functions are written down and some issues are discussed. In section \ref{tcr} I discuss different approaches aimed at solving the equations at three-level, i.e. in the large-$N$ limit. Finally, section \ref{outlook} concludes this paper with an outlook at future possible avenues of research.

\section{Schwinger--Dyson on the lattice} \label{sdotl}

\subsection{The lattice functional derivative}
In order to write down Schwinger--Dyson equations, one needs to define a derivation operator Stokes' theorem can be applied to. On the lattice, the integrations are integrations over the gauge group, and a derivative exists having the desired properties \cite{Kerler:1980ke}:
\begin{equation}
\delta^a_U Q(U,\ldots) = \lim_{\epsilon\to0} \frac{\mathcal Q(e^{i\tau^a/2}U,\ldots) - \mathcal Q(U,\ldots)}\epsilon \;,
\end{equation}
where $\tau^a$ are the generators of the gauge group under question, obeying the commutation relations $[\tau^a,\tau^b]=2if^{abc}\tau^c$ with $f^{abc}$ the structure constants, and the dots denote other stuff the quantity $\mathcal Q$ may depend on. Mark that the derivative carries a gauge group index $a$. When computing the integral of such a derivative, the result will be zero due to the invariance of the integration measure under $U\to VU$ with $V$ an arbitrary gauge group element. A right derivative can be defined in a completely analogous way.

In practice, this functional derivative operates as
\begin{equation}
\delta_U^a U = \frac{i\tau^a}2 U \;, \qquad \delta_U^a U^\dagger = U^\dagger \left(-\frac{i\tau^a}2\right) \;,
\end{equation}
and it furthermore obeys Leibniz's rule.\footnote{The derivative operator itself commutes with all other quantities.} When working on a Wilson loop containing the link matrix $U_{x,\mu}$, the derivative will ``cut open'' the loop and introduce a group generator. One can depict this as:
\begin{equation}
\delta^a_{x,\mu} \quad
\parbox{3\unitlength}{\begin{picture}(3,3)
\thicklines
\put(0,3){\line(0,-1){3}} \put(0,0){\vector(1,0){2}} \put(2,0){\line(1,0){1}} \put(3,0){\line(0,1){3}} \put(-.7,-.7){$x$} \put(1.5,-1){$\hat\mu$} \put(2.2,-.8){\vector(1,0){1}}
\end{picture}}
\quad = \quad
\parbox{3\unitlength}{\begin{picture}(3,3)
\thicklines
\put(0,3){\line(0,-1){3}} \put(0,0){\vector(1,0){2}} \put(2,0){\line(1,0){1}} \put(3,0){\line(0,1){3}} \put(0,0){\circle*{.4}} \put(.5,.5){$a$} \put(-.7,-.7){$x$} \put(1.5,-1){$\hat\mu$} \put(2.2,-.8){\vector(1,0){1}}
\end{picture}}
\quad , \qquad
\delta^a_{x,\mu} \quad
\parbox{3\unitlength}{\begin{picture}(3,3)
\thicklines
\put(0,3){\line(0,-1){3}} \put(3,0){\vector(-1,0){2}} \put(0,0){\line(1,0){1}} \put(3,0){\line(0,1){3}} \put(-.7,-.7){$x$} \put(1.5,-1){$\hat\mu$} \put(2.2,-.8){\vector(1,0){1}}
\end{picture}}
\quad = - \quad
\parbox{3\unitlength}{\begin{picture}(3,3)
\thicklines
\put(0,3){\line(0,-1){3}} \put(3,0){\vector(-1,0){2}} \put(0,0){\line(1,0){1}} \put(3,0){\line(0,1){3}} \put(0,0){\circle*{.4}} \put(.5,.5){$a$} \put(-.7,-.7){$x$} \put(1.5,-1){$\hat\mu$} \put(2.2,-.8){\vector(1,0){1}}
\end{picture}} \quad .
\end{equation} \vspace{.5\unitlength} \\
Here, the arrow shows the orientation of the Wilson loop (i.e.: the loop in the first identity contains the matrix $U_{x,\mu}$, the one in the second identity contains $U_{x,\mu}^\dagger$). The dot with index $a$ has been added at the place where the group generator $i\tau^a/2$ has been inserted. A derivative $\delta_{x+\hat\mu,-\mu}$,\footnote{I will use the notation $\hat\mu$ for the vector with length the lattice spacing in the direction $\mu$ throughout.} working on the same link matrix but coming from the other side, gives an analogous result but with the signs flipped around.

In order to find a gauge invariant result from the lattice functional derivative, one needs a way to close the Wilson loops that have been cut open by the derivative. Given the well-known identity for the generators of the gauge group
\begin{equation} \label{switchtau}
\sum_a \tau^a_{\alpha\beta}\tau^a_{\gamma\delta} = 2 \delta_{\alpha\delta}\delta_{\beta\gamma} - \frac2N \delta_{\alpha\beta}\delta_{\gamma\delta} \;,
\end{equation}
where $N$ is the number of colors (in SU($N$)), it is clear that having two generator insertions at the same lattice site and summing over the color index will close the loops:
\begin{equation} \label{switching}
\sum_a
\parbox{6em}{\begin{picture}(6,6)
\thicklines
\put(0,3.2){\vector(1,0){2}} \put(2,3.2){\line(1,0){.8}} \put(2.8,3.2){\circle*{.4}} \put(2.8,3.2){\line(0,1){2.8}} \put(2.1,3.6){$a$}
\put(6,2.8){\vector(-1,0){2}} \put(4,2.8){\line(-1,0){.8}} \put(3.2,2.8){\circle*{.4}} \put(3.2,2.8){\line(0,-1){2.8}} \put(3.6,2.1){$a$}
\put(3.2,3.2){$x$}
\end{picture}}
\quad = \frac1{2N} \quad
\parbox{6em}{\begin{picture}(6,6)
\thicklines
\put(0,3.2){\vector(1,0){2}} \put(2,3.2){\line(1,0){.8}} \put(2.8,3.2){\line(0,1){2.8}}
\put(6,2.8){\vector(-1,0){2}} \put(4,2.8){\line(-1,0){.8}}\put(3.2,2.8){\line(0,-1){2.8}}
\put(3.2,3.2){$x$}
\end{picture}}
\quad - \frac12 \quad
\parbox{6em}{\begin{picture}(6,6)
\thicklines
\put(0,2.8){\vector(1,0){2}} \put(2,2.8){\line(1,0){.8}} \put(2.8,2.8){\line(0,-1){2.8}}
\put(6,3.2){\vector(-1,0){2}} \put(4,3.2){\line(-1,0){.8}} \put(3.2,3.2){\line(0,1){2.8}}
\put(2.4,3.2){$x$}
\end{picture}} \quad .
\end{equation}
One sees that, in order to have closed loops as a result, it suffices to already have some generator with same gauge index $a$ present at the lattice site where the derivative is going to be taken, after which one has but to sum over the color index.

When the derivative in this gauge invariant combination hits the piece of Wilson loop where the generator matrix was inserted, another useful identity can be used:
\begin{equation} \label{tweedeafgeleide}
\sum_a \delta^a_{x,\mu} \frac{i\tau^a}2 U_{x,\mu} = -\frac{N^2-1}{2N} U_{x,\mu} \;.
\end{equation}
This is an immediate result of the identity \eqref{switchtau}.

\subsection{Towards the equations}
All this together bring us to the strategy to follow for writing down Schwinger--Dyson equations. The master equation is:\footnote{In the typical way of constructing Schwinger--Dyson equations, one introduces a source term, and different equations are obtained by taking derivative with respect to the source. In the case at hand it seems more convenient to put the relevant fields already in the path integral instead, as one of the loops should contain a gauge group generator.}
\begin{equation} \label{master}
\sum_a \int [dU] \delta^a_{x,\mu} (\mathcal Q_x[U] \mathcal R[U] e^{-S[U]}) = 0 \;.
\end{equation}
Here, $\mathcal Q_x[U]$ is some Wilson loop that has a gauge group generator insertion at the site $x$, $\mathcal R[U]$ is a product of some Wilson loops (or none), and $S[U]$ is the Yang--Mills action. In this paper, the action will be taken to be the ordinary Wilson action:
\begin{equation}
S = \frac N{g^2} \sum_{\substack{x\\\mu\neq\nu}} \left(1-\frac1N P_{\mu\nu}(x)\right) \;,
\end{equation}
where $P_{\mu\nu}(x)$ is the one-plaquette Wilson loop in the $(\hat\mu,\hat\nu)$ plane centered on $x$.

A first term encountered when working out \eqref{master} is the one where the derivative hits $\mathcal Q_x[U]$ in the link where the gauge group generator insertion has been placed. The identity \eqref{tweedeafgeleide} will be triggered, and the resulting term is nothing but
\begin{equation} \label{eerstetermniets}
-\frac{N^2-1}{2N} \langle \mathcal Q[U] \mathcal R[U] \rangle \;.
\end{equation}
A second contribution comes from the derivative hitting the loop $\mathcal Q$ in another instance of the $U_{x,\mu}$ matrix, as can happen if the loop doubles back on itself. A first term from this is the one coming from the first term in \eqref{switching}, where the generators are just deleted and a contribution proportional to \eqref{eerstetermniets} arises. An additional term comes from the switching around of the two pieces of loop, which will result in the loop $\mathcal Q$ falling apart into two sub-loops. As an example, consider a Wilson loop winding twice around an elementary plaquette, with one insertion of a generator and a derivative at the same site:
\begin{equation} \label{dubbelvaltuitelkaar}
\sum_a \delta^a_{x_0,x} \; \parbox{4em}{\begin{picture}(4,3)
\thicklines
\put(0,3){\line(0,-1){2.6}} \put(.4,0){\vector(1,0){1.6}} \put(2,0){\line(1,0){1}} \put(3,0){\line(0,1){3}} \put(0,3){\line(1,0){3}} \put(0,.4){\line(1,0){2.6}} \put(2.6,.4){\line(0,1){2.2}} \put(2.6,2.6){\line(-1,0){2.2}} \put(.4,2.6){\line(0,-1){2.6}} \put(.4,0){\circle*{.4}} \put(-.8,-.2){$x_0$} \put(0,-.9){$a$}
\end{picture}}
= \sum_a \left( \ \ \parbox{4em}{\begin{picture}(4,3)
\thicklines
\put(0,3){\line(0,-1){2.6}} \put(.4,0){\vector(1,0){1.6}} \put(2,0){\line(1,0){1}} \put(3,0){\line(0,1){3}} \put(0,3){\line(1,0){3}} \put(0,.4){\line(1,0){2.6}} \put(2.6,.4){\line(0,1){2.2}} \put(2.6,2.6){\line(-1,0){2.2}} \put(.4,2.6){\line(0,-1){2.6}} \put(.4,0){\circle*{.4}} \put(-.4,-.2){$a$} \put(.8,0){\circle*{.4}} \put(.6,-.9){$a$}
\end{picture}}
+ \quad \parbox{4em}{\begin{picture}(4,3)
\thicklines
\put(0,3){\line(0,-1){2.6}} \put(.4,0){\vector(1,0){1.6}} \put(2,0){\line(1,0){1}} \put(3,0){\line(0,1){3}} \put(0,3){\line(1,0){3}} \put(0,.4){\line(1,0){2.6}} \put(2.6,.4){\line(0,1){2.2}} \put(2.6,2.6){\line(-1,0){2.2}} \put(.4,2.6){\line(0,-1){2.6}} \put(.4,0){\circle*{.4}} \put(0,-.9){$a$} \put(0,.4){\circle*{.4}} \put(-.7,0){$a$}
\end{picture}} \right)
= -\frac{N^2-2}{2N} \; \parbox{4em}{\begin{picture}(4,3)
\thicklines
\put(0,3){\line(0,-1){2.6}} \put(.4,0){\vector(1,0){1.6}} \put(2,0){\line(1,0){1}} \put(3,0){\line(0,1){3}} \put(0,3){\line(1,0){3}} \put(0,.4){\line(1,0){2.6}} \put(2.6,.4){\line(0,1){2.2}} \put(2.6,2.6){\line(-1,0){2.2}} \put(.4,2.6){\line(0,-1){2.6}}
\end{picture}}
- \frac12 \ \ \parbox{4em}{\begin{picture}(4,3)
\thicklines
\put(0,3){\line(0,-1){3}} \put(0,0){\vector(1,0){2}} \put(2,0){\line(1,0){1}} \put(3,0){\line(0,1){3}} \put(0,3){\line(1,0){3}} \put(.4,.4){\vector(1,0){1.8}} \put(.4,.4){\line(1,0){2.2}} \put(2.6,.4){\line(0,1){2.2}} \put(2.6,2.6){\line(-1,0){2.2}} \put(.4,2.6){\line(0,-1){2.2}}
\end{picture}} . \parbox[t][1.8em][t]{0em}{~}
\end{equation}
Here, the $x$-axis is taken to point to the right.

If, in the master equation \eqref{master}, one of the loops in $\mathcal R[U]$ contains the link the derivative is working on, there will be additional terms from this. A last contribution comes from elementary plaquette Wilson loops coming from the action in the exponential. These will give higher $n$-point functions containing these elementary plaquettes as well as terms wherein the loop $\mathcal Q$ has been augmented with these plaquettes.

\section{Constructing the equations} \label{cte}

\subsection{One-point functions}
\begin{figure} \centering
\subfloat[$P_{\mu\nu}(x)$]{\begin{picture}(6,6)
\thicklines
\put(2,5){\line(0,-1){3}} \put(2,2){\vector(1,0){2}} \put(4,2){\line(1,0){1}} \put(5,2){\line(0,1){3}} \put(2,5){\line(1,0){3}} \put(3.5,3.5){\circle*{.4}} \put(4,3.2){$x$} \put(3,.7){\vector(1,0){3}} \put(3.5,.0){$\mu$} \put(.7,3){\vector(0,1){3}} \put(0,3.5){$\nu$}
\end{picture}
\label{plaquette}}
\qquad\qquad
\subfloat[$D_{\mu\nu}(x)$ \label{dubbel}]{\begin{picture}(6,6)
\thicklines
\put(2,5){\line(0,-1){2.6}} \put(2.4,2){\vector(1,0){1.6}} \put(4,2){\line(1,0){1}} \put(5,2){\line(0,1){3}} \put(2,5){\line(1,0){3}} \put(2,2.4){\line(1,0){2.6}} \put(4.6,2.4){\line(0,1){2.2}} \put(4.6,4.6){\line(-1,0){2.2}} \put(2.4,4.6){\line(0,-1){2.6}} \put(3.5,3.5){\circle*{.4}} \put(4,3.2){$x$} \put(3,.7){\vector(1,0){3}} \put(3.5,0){$\mu$} \put(.7,3){\vector(0,1){3}} \put(0,3.5){$\nu$}
\end{picture}}
\qquad\qquad
\subfloat[$S_{\mu\nu\lambda}(x)$]{\begin{picture}(9,7.5)
\thicklines
\put(2,3.5){\line(0,1){3}} \put(2,6.5){\line(1,0){1}} \put(5,6.5){\vector(-1,0){2}} \put(5,3.5){\line(0,1){3}} \put(2,3.5){\line(4,-3){2}} \put(4,2){\line(1,0){3}} \put(5,3.5){\line(4,-3){2}} \put(.7,4.5){\vector(0,1){3}} \put(2,2){\vector(4,-3){2}} \put(6,3.5){\vector(1,0){3}} \put(3.5,3.5){\circle*{.4}} \put(0,5){$\nu$} \put(7.5,2.8){$\mu$} \put(2,.5){$\lambda$} \thinlines \multiput(2,3.5)(.6,0){5}{\line(1,0){.4}} \put(3.5,3.8){$x$}
\end{picture}} \\
\subfloat[$R_{\mu\nu}(x)$]{\begin{picture}(6,9)
\thicklines
\put(2,8){\line(0,-1){6}} \put(2,2){\vector(1,0){2}} \put(4,2){\line(1,0){1}} \put(5,2){\line(0,1){6}} \put(2,8){\line(1,0){3}} \put(3.5,5){\circle*{.4}} \put(3.5,5.3){$x$} \put(3,.7){\vector(1,0){3}} \put(3.5,0){$\mu$} \put(.7,6){\vector(0,1){3}} \put(0,6.5){$\nu$} \thinlines \multiput(2,5)(.6,0){5}{\line(1,0){.4}}
\end{picture}}
\qquad\qquad
\subfloat[$F_{\mu\nu}(x)$]{\begin{picture}(6,9)
\thicklines
\put(2,8){\line(0,-1){2.8}} \put(2,2){\line(0,1){2.8}} \put(4.6,2){\vector(-1,0){2}} \put(2,2){\line(1,0){1}} \put(4.6,2){\line(0,1){3.2}} \put(5,4.8){\line(0,1){3.2}} \put(2,8){\line(1,0){3}} \put(3.5,5){\circle*{.4}} \put(3.5,5.5){$x$} \put(3,.7){\vector(1,0){3}} \put(3.5,0){$\mu$} \put(.7,6){\vector(0,1){3}} \put(0,6.5){$\nu$} \put(2,4.8){\line(1,0){3}} \put(2,5.2){\line(1,0){2.6}} \put(5,8){\vector(-1,0){2}}
\end{picture}}
\qquad\qquad
\subfloat[$C_{\mu\nu\lambda}(x)$]{\begin{picture}(9,7.5)
\thicklines
\put(2,3.7){\line(0,1){2.8}} \put(2,6.5){\line(1,0){1}} \put(5,6.5){\vector(-1,0){2}} \put(5,3.3){\line(0,1){3.2}} \put(2,3.3){\line(4,-3){2}} \put(4,1.8){\line(1,0){3}} \put(4.6,3.7){\line(4,-3){2.4}} \put(.7,4.5){\vector(0,1){3}} \put(2,2){\vector(4,-3){2}} \put(6,3.5){\vector(1,0){3}} \put(3.5,3.5){\circle*{.4}} \put(0,5){$\nu$} \put(7.5,2.8){$\mu$} \put(2,.5){$\lambda$} \put(3.5,4){$x$} \put(2,3.7){\line(1,0){2.7}} \put(2,3.3){\line(1,0){3}}
\end{picture}}
\caption{The one-plaquette operator $P$ and the five two-plaquette operators: the double plaquette $D$, the seat $S$, the rectangle $R$, the flipped rectangle $F$ and the folding chair $C$.}
\label{operatoren}
\end{figure}
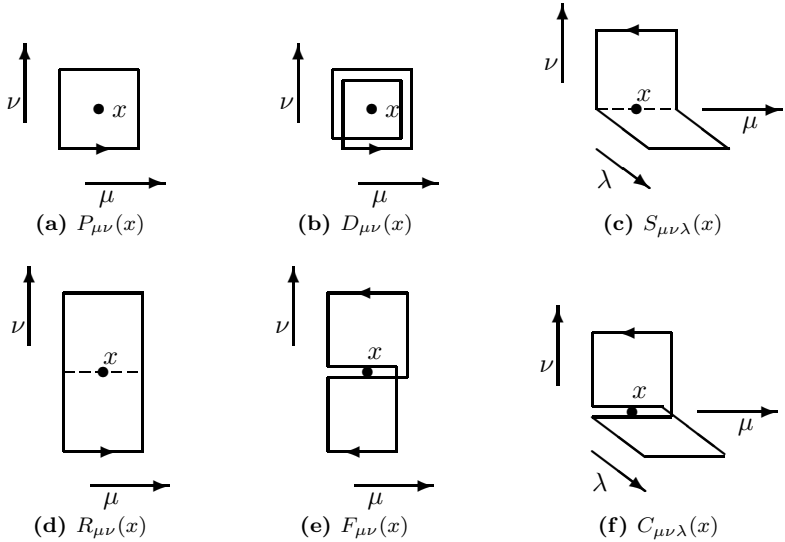

As a first case, let us consider the equations for vacuum expectation values of Wilson loops. These have already been derived in the literature, but for completeness I will repeat the derivation using the notations introduced in this paper. First let me introduce a slightly deviant convention which will simplify the look of the equations later on. With $P_{\mu\nu}(x)$ I will denote the Wilson plaquette centered at $x$ turning in positive direction in the coordinate system $(\hat\mu,\hat\nu)$. This is depicted in \figurename\ \ref{plaquette}. The vacuum expectation value of the plaquette divided by the number of colors will be denoted $P$. Mark that an operator like $P_{\nu\mu}(x)$ is equal to $P_{\mu\nu}^*(x)$.

There are some more operators needed in this paragraph. They are also drawn in \figurename\ \ref{operatoren}. For these operators I will again use the convention of indexless letters to denote the vacuum expectation values, thus: $D = \langle \frac1N D_{\mu\nu}(x)\rangle$ etc. Furthermore, it will now and then be useful to allow indices to take negative values, with which I mean that the direction in question is taken in opposite sense. This means that, for example, $P_{-\mu\nu}(x) = P_{\mu\nu}^*(x)$.

\subsubsection{The elementary plaquette}
As a first example, let us consider the master formula \eqref{master} with $\mathcal Q[U]$ the elementary plaquette Wilson loop $P$ centered at the origin, $\mathcal R[U]$ absent, and the derivative taken along one of the links of the loop under consideration. For concreteness, let the loop be in the $(x,y)$ plane, with the derivative working on the site at $-\frac12(\hat x+\hat y)$ along the $x$-axis. The starting point is:
\begin{equation}
\sum_a \int [dU] \delta^a_x(-\tfrac{\hat x+\hat y}2) \left(P_{xy}^{-(\hat x+\hat y)/2}(0) e^{-S}\right) = 0 \;.
\end{equation}
The superscript at the plaquette indicates that it has been cut open with a group generator at that lattice site.

The first term in the equation is the one like in \eqref{eerstetermniets}:
\begin{equation}
-\frac{N^2-1}2 P \;,
\end{equation}
where one factor of $N$ has been absorbed in the vacuum expectation value of the loop.

The second contribution comes from elementary loops coming down from the action. Firstly, there is the term where the loops are not switched around (i.e. the contributions from the first term in \eqref{switching}), which introduces all elementary loops containing the link of the functional derivative, with a minus sign if the loop flows in opposite direction from the derivative:
\begin{equation}
\frac1{2g^2N} \left\langle P_{xy}(0) \sum_{\pm\mu\neq x} \left(P_{x\mu}(\tfrac{\hat \mu-\hat y}2) - P_{\mu x}(\tfrac{\hat \mu-\hat y}2)\right) \right\rangle \;.
\end{equation}
The sum over $\mu$ is here over both positive and negative senses, in all directions except along the $x$ axis. If now we introduce the propagator notation
\begin{equation}
\Delta_{\mu\nu,\kappa\lambda}(x-y) = \langle P_{\mu\nu}(x) P_{\lambda\kappa}(y)\rangle_\text{conn} = \langle P_{\mu\nu}(x) P_{\lambda\kappa}(y)\rangle - N^2P^2 \;,
\end{equation}
the above result can be rewritten as
\begin{equation}
\frac1{2g^2N} \sum_{\pm\mu\neq x} \left(\Delta_{xy,x\mu}(\tfrac{\hat\mu-\hat y}2) - \Delta_{xy,\mu x}(\tfrac{\hat\mu-\hat y}2)\right) \;.
\end{equation}

Then there is the part where the loops coming from the action and the original loop put in the path integral are linked with each other, through the second term in \eqref{switching}, giving
\begin{multline}
-\frac1{2g^2} \int [dU] \left(D_{xy}(0) - N + F_{xy}(-\tfrac{\hat y}2) - R_{xy}(-\tfrac{\hat y}2) + \sum_{\pm\mu\neq x,y} \left(C_{xy\mu}(-\tfrac{\hat y}2) - S_{xy\mu}(-\tfrac{\hat y}2)\right)\right) e^{-S} \\
= -\frac N{2g^2} (D-1+F-R+2(d-2)(C-S))
\end{multline}
where $d$ is the spacetime dimensionality. The term of $N$ comes from a $\tr\mathbbm1$, caused by a loop being reconnected to its mirror image.

All together we find the following equation:
\begin{multline} \label{eenplaketresultaat}
-\frac{N^2-1}2P + \frac1{2g^2N} (\Delta_D - \Delta_{\mathbbm1} + \Delta_F - \Delta_R + 2(d-2)(\Delta_C-\Delta_S)) \\ - \frac N{2g^2} (D-1+F-R+2(d-2)(C-S)) = 0 \;,
\end{multline}
where $\Delta_{\mathcal O}$ is a shorthand for the propagator between two plaquettes that share at least one link, and which are positioned such that linking them with each other would give the operator $\mathcal O$, as for example $\Delta_D = \Delta_{xy,xy}(0)$. This gives a relation between the values of certain condensates and the propagator of the operator $P_{\mu\nu}$ at infinitesimal (lattice-spacing) distance.

In order to make contact with the large-$N$ expansion, let us introduce the 't~Hooft coupling $\lambda=g^2N$. Using this, the above equation can be rewritten as
\begin{multline}
\left(1-\frac1{N^2}\right) P + \frac1\lambda (D+F-R+2(d-2)(C-S)) = \frac1\lambda  \\ + \frac1{N^2} \frac1\lambda (\Delta_D - \Delta_{\mathbbm1} + \Delta_F - \Delta_R + 2(d-2)(\Delta_C-\Delta_S)) \;.
\end{multline}
This equation can be schematically depicted as
\begin{equation} \label{eenpuntvgl}
\parbox{4em}{\begin{fmffile}{eenpunt} \begin{fmfgraph*}(3,2) \fmfleft{l} \fmfright{r} \fmf{plain}{l,r} \fmfv{decor.shape=circle,decor.filled=full,decor.size=5thick}{r} \end{fmfgraph*} \end{fmffile}} + \ \parbox{4em}{\begin{fmffile}{eenpuntQ} \begin{fmfgraph*}(3,2) \fmfleft{l} \fmfright{r} \fmf{dbl_plain}{l,r} \fmfv{decor.shape=circle,decor.filled=full,decor.size=5thick}{r} \end{fmfgraph*} \end{fmffile}} = \parbox{4em}{\begin{fmffile}{eenpunttree} \begin{fmfgraph*}(3,2) \fmfleft{l} \fmfright{r} \fmf{plain}{l,r} \fmfdot{r} \end{fmfgraph*} \end{fmffile}} + \frac1{N^2} \parbox{5em}{\begin{fmffile}{eenpuntloop} \begin{fmfgraph*}(3,2) \fmfleft{l1,l2} \fmfright{r} \fmf{phantom}{l1,v,l2} \fmf{plain,tension=.2}{v,v} \fmfiv{decor.shape=circle,decor.filled=full,decor.size=5thick}{(17.5thick,5thick)} \end{fmfgraph*} \end{fmffile}}.
\end{equation}
The left-hand side contains the one-plaquette condensate (single line) together with several other condensates (double line), which are the unknowns of the equation. At the right-hand side one has a tree-level value for the condensates, and a loop-correction which is suppressed by a coupling constant equal to $1/N^2$. Mark that the first coefficient at the left-hand side receives a $1/N^2$ correction which does not originate from any loop-like diagram, but which is rather similar to a finite renormalization.

\subsubsection{Other Wilson loops}
Equations for other one-point functions can be written down in a completely analogous way. Several issues immediately arise.

When writing down the equations for, for example, the double plaquette $D$ condensate (see \figurename\ \ref{dubbel}), it will contain a nonlinear term due to the recombination coming from the second contribution shown in \eqref{dubbelvaltuitelkaar}. The presence of such nonlinearities in the condensate equations seriously complicate solving them.

Another point of note is the fact that there exists a certain arbitrariness in the construction of the equations. In order to write down the equation for a certain condensate, a site and link in the relevant Wilson loop have to be chosen, and a different choice will, of course, lead to a different equation. The solution must still be the same, which it can be verified to be in those cases where explicit expressions for (parts of) the solution can be obtained, like in the limit of large 't~Hooft coupling $\lambda$ within the large-$N$ limit. In the following I will systematically symmetrize the choice of site and link over the entire loop under consideration, so as to get more symmetric equations. What extra information, if any, can be gained from making different choices is still to be explored.

\subsection{Two-point functions}
Let us now consider the Schwinger--Dyson equations obtained by putting two one-plaquette operators into the path integral. One of these plaquettes is positioned at the origin as in the one-plaquette example, and the second one has an arbitrary position $x$. The expression we consider is
\begin{equation} \label{sdeendelta}
\frac14 \sum_{\text{links } l} \sum_a \int [dU] \delta^a_l \left(P_{\alpha\beta}^{a,l}(0) P_{\gamma\delta}(x) e^{-S}\right) = 0 \;.
\end{equation}
The $\frac14\sum_l$ is a symmetrization over the four links of the square plaquette at the origin, which will make the equations manifestly symmetric under rotation over a right angle.\footnote{Unsymmetrized equations must also be fulfilled by an eventual solution of the theory, but, pending the development of the necessary techniques to find such a solution, it seems one gets most physical intuition from the symmetrized equations.}

The equations can be straightforwardly written down following the same procedure as in the previous section. The main difference is the extra contribution arising whenever the second plaquette $P_{\gamma\delta}(x)$ has some link in common with the first one. In that case, two extra term appear due to a functional derivative working on this second plaquette, giving one term where the plaquettes are left unchanged and one term with the two Wilson loops merged into one. Schematically, the result looks like:\footnote{The full expression is quite lenghty, especially for general indices and positions, but the interested reader should not encounter too many difficulties in deriving it for himself.}
\begin{equation} \label{tweepuntvgl}
\parbox{5.2em}{\begin{fmffile}{tweepunt} \begin{fmfgraph*}(5,2) \fmfkeep{tweepunt} \fmfleft{l} \fmfright{r} \fmf{plain}{l,v,r} \fmfv{decor.shape=circle,decor.filled=full,decor.size=5thick}{v} \end{fmfgraph*} \end{fmffile}} + \ \parbox{5.2em}{\begin{fmffile}{tweepuntQ} \begin{fmfgraph*}(5,2) \fmfkeep{tweepuntQ} \fmfleft{l} \fmfright{r} \fmf{dbl_plain}{l,v} \fmf{plain}{v,r} \fmfv{decor.shape=circle,decor.filled=full,decor.size=5thick}{v} \end{fmfgraph*} \end{fmffile}} = N^2 \quad \parbox{3.2em}{\begin{fmffile}{tweepunteen} \begin{fmfgraph*}(3,2) \fmfleft{l} \fmfright{r} \fmf{plain}{l,r} \fmfv{decor.shape=circle,decor.filled=full,decor.size=5thick}{l} \end{fmfgraph*} \end{fmffile}} + \delta(x) \left(\alpha + \parbox{1em}{\begin{fmffile}{tweepuntcond} \begin{fmfgraph*}(1,2) \fmfkeep{tweepuntcond} \fmftop{t} \fmfbottom{b} \fmf{dbl_plain}{t,b} \fmfv{decor.shape=circle,decor.filled=full,decor.size=5thick}{t} \end{fmfgraph*} \end{fmffile}} + \frac1{N^2} \parbox{2em}{\begin{fmffile}{tweepuntprop} \begin{fmfgraph*}(2,2) \fmfkeep{tweepuntprop} \fmfbottom{b1,b2} \fmf{phantom}{b1,v,b2} \fmf{plain,tension=.3}{v,v} \fmfiv{decor.shape=circle,decor.filled=full,decor.size=5thick}{(5thick,12.5thick)} \end{fmfgraph*} \end{fmffile} \vspace{-1.5em}}\right) + \frac1N \ \parbox{6em}{\begin{fmffile}{tweepuntloop} \begin{fmfgraph*}(6,2) \fmfleft{l} \fmfright{r} \fmf{plain}{l,v1} \fmf{plain,tension=.5}{v2,r} \fmf{plain,left,tension=.1}{v1,v2,v1} \fmfv{decor.shape=circle,decor.filled=full,decor.size=5thick}{v2} \end{fmfgraph*} \end{fmffile}} \ .
\end{equation}
All $n$-point functions with blobs denote full $n$-point function, which contain the disconnected parts. In the above equation, $\delta(x)$ depicts schematically the Kronecker deltas coming from all different configurations where the two plaquettes touch each other. The constant $\alpha$ is one if the two plaquettes in the equation exactly cover each other but have opposite orientation, and zero otherwise.

The next step is to separate all $n$-point in a connected and a disconnected part. The disconnected parts coming from all terms --- except the one with the Kronecker delta --- cancel due to the equation for the one-point function. Next, we go to Fourier space using the transforms
\begin{equation}
\tilde f(p) = \sum_x e^{ix\cdot p} f(x) \;, \qquad f(x) = \int_{-\pi}^{+\pi} e^{-ix\cdot p} \tilde f(p) \frac{d^dp}{(2\pi)^d} \;,
\end{equation}
where the integral in the second identity is over a hypercubic box of size $2\pi$. The result is a system of linear equations for the propagators $\tilde\Delta_{\alpha\beta,\gamma\delta}(p)$. It is possible to decouple these equations according to charge conjugation by taking symmetric and antisymmetric linear combinations of the Lorentz indices. Defining $P_{\mu\nu}^\pm = \frac12(P_{\mu\nu} \pm P_{\nu\mu})$ (which are nothing but the real and imaginary parts of the plaquette operator $P_{\mu\nu}$), one gets the new propagators
\begin{equation}
\Delta_{\mu\nu,\alpha\beta}^{st}(x) = \left\langle P_{\mu\nu}^s(0) P_{\alpha\beta}^t(x)\right\rangle_\text{conn} \;.
\end{equation}
With these definitions, there are four propagators, but the ones mixing different charge conjugation states must be zero due to the charge conjugation symmetry. This turns out to be consistent with the equations. For the two nonzero propagators we get:
\begin{subequations} \label{tweeprops}
\begin{multline} \label{plusplus}
\left(1-\frac1{N^2}\right) \tilde\Delta_{\alpha\beta,\gamma\delta}^{++}(p)
= \frac{\delta_{\alpha\gamma}\delta_{\beta\delta} + \delta_{\alpha\delta}\delta_{\beta\gamma}}2 \Bigg(1 - D + F - R + \frac{\Delta_D-\Delta_{\mathbbm1} + \Delta_R-\Delta_F}{N^2} \\ - (\cos^2\tfrac{p_\alpha}2+\cos^2\tfrac{p_\beta}2) \left(F - R + S - C + \frac{\Delta_R-\Delta_F + \Delta_C-\Delta_S}{N^2}\right)\Bigg) \\
+ C^{++}_{\alpha\beta,\gamma\delta}(p) \left(S - C + \frac{\Delta_C-\Delta_S}{N^2}\right)
- \frac1{4\lambda} \tilde\Delta_{\alpha\beta,\gamma\delta}^{QP,++}(p) \\
- \frac2{N\lambda} \int_{-\pi}^{+\pi} dq \left(\sum_{\mu\neq\alpha} \sin\tfrac{q_\mu}2 \sin\tfrac{q_\beta}2 \tilde\Gamma_{\alpha\beta,\alpha\mu,\gamma\delta}^{--+}(-q-p,q,p) + \sum_{\lambda\neq\beta} \sin\tfrac{q_\mu}2 \sin\tfrac{q_\alpha}2 \tilde\Gamma_{\alpha\beta,\mu\beta,\gamma\delta}^{--+}(-q-p,q,p)\right)
\end{multline}
and
\begin{multline} \label{minmin}
\left(1-\frac1{N^2}\right) \tilde\Delta_{\alpha\beta,\gamma\delta}^{--}(p) - \frac{2P}\lambda \left(\sum_{\mu\neq\alpha} \tilde\Delta_{\alpha\mu,\gamma\delta}^{--}(p) \sin\tfrac{p_\mu}2 \sin\tfrac{p_\beta}2 + \sum_{\mu\neq\beta} \sin\tfrac{p_\alpha}2 \sin\tfrac{p_\mu}2 \tilde\Delta_{\mu\beta,\gamma\delta}^{--}\right) \\
= - \frac{\delta_{\alpha\gamma}\delta_{\beta\delta} - \delta_{\alpha\delta}\delta_{\beta\gamma}}2 \Bigg(1 + D - F - R - \frac{\Delta_D+\Delta_{\mathbbm1}-\Delta_R-\Delta_F}{N^2} \\ + (\sin^2\tfrac{p_\alpha}2 + \sin^2\tfrac{p_\beta}2) \left(F + R - S - C - \frac{\Delta_R+\Delta_F-\Delta_C-\Delta_S}{N^2}\right)\Bigg) \\
+ S^{--}_{\alpha\beta,\gamma\delta}(p) \left(2P^2 - S - C + \frac{\Delta_C+\Delta_S}{N^2}\right)
- \frac1{4\lambda} \tilde\Delta_{\alpha\beta,\gamma\delta}^{QP,--}(p) \\
- \frac2{N\lambda} \int_{-\pi}^{+\pi} dq \left(\sum_{\mu\neq\alpha} \sin\tfrac{q_\mu}2 \sin\tfrac{q_\beta}2 \tilde\Gamma_{\alpha\beta,\alpha\mu,\gamma\delta}^{+--}(-q-p,q,p) + \sum_{\mu\neq\beta} \sin\tfrac{q_\mu}2 \sin\tfrac{q_\alpha}2 \tilde\Gamma_{\alpha\beta,\mu\beta,\gamma\delta}^{+--}(-q-p,q,p)\right) \;.
\end{multline}
\end{subequations}
where the following shorthands have been used:
\begin{subequations} \begin{gather}
C^{++}_{\alpha\beta,\gamma\delta}(p) = \frac{\delta_{\alpha\gamma}\cos\tfrac{p_\beta}2\cos\tfrac{p_\delta}2 + \delta_{\beta\gamma}\cos\tfrac{p_\alpha}2\cos\tfrac{p_\delta}2 + \delta_{\alpha\delta}\cos\tfrac{p_\beta}2\cos\tfrac{p_\gamma}2 + \delta_{\beta\delta}\cos\tfrac{p_\alpha}2\cos\tfrac{p_\gamma}2}2 \\
S^{--}_{\alpha\beta,\gamma\delta}(p) = \frac{\delta_{\alpha\gamma}\sin\tfrac{p_\beta}2\sin\tfrac{p_\delta}2 - \delta_{\beta\gamma}\sin\tfrac{p_\alpha}2\sin\tfrac{p_\delta}2 - \delta_{\alpha\delta}\sin\tfrac{p_\beta}2\sin\tfrac{p_\gamma}2 + \delta_{\beta\delta}\sin\tfrac{p_\alpha}2\sin\tfrac{p_\gamma}2}2
\end{gather} \end{subequations}
and where furthermore the operator $Q_{\mu\nu}(x)$ has been defined as four times the $D_{\mu\nu}(x)$ operator, plus the sum of all $F$ and $C$ operators containing $P_{\mu\nu}(x)$, minus all the $R$ and $S$ operator containing this same plaquette. Furthermore, the connected three-point function has been introduced as
\begin{equation}
\Gamma_{\alpha\beta,\gamma\delta,\epsilon\zeta}(x,y,z) = \left\langle P_{\alpha\beta}(x) P_{\gamma\delta}(y) P_{\epsilon\zeta}(z) \right\rangle_\text{conn} \;,
\end{equation}
and with similar expressions for symmetrized and antisymmetrized three-point functions. Its Fourier transform has been defined as
\begin{equation}
\tilde\Gamma_{\alpha\beta,\gamma\delta,\epsilon\zeta}(-p-q,p,q) = \sum_{x,y} e^{i(x\cdot p+y\cdot q)} \Gamma_{\alpha\beta,\gamma\delta,\epsilon\zeta}(0,x,y) \;.
\end{equation}

The above equations can be depicted as
\begin{equation}
\mx{X} \parbox{5.2em}{\fmfreuse{tweepunt}} + \ \parbox{5.2em}{\fmfreuse{tweepuntQ}} = \mx{Y} + \frac1N \ \parbox{8em}{\begin{fmffile}{proploop} \begin{fmfgraph*}(8,4) \fmfleft{l} \fmfright{r} \fmf{plain,tension=2}{l,v1} \fmf{plain,tension=1.3}{v2,v3,r} \fmf{plain,left,tension=.3}{v1,v2,v1} \fmfv{decor.shape=circle,decor.filled=full,decor.size=5thick}{v2} \fmfv{decor.shape=circle,decor.filled=full,decor.size=5thick}{v3} \fmfiv{decor.shape=circle,decor.filled=full,decor.size=5thick}{(12.5thick,17.5thick)} \fmfiv{decor.shape=circle,decor.filled=full,decor.size=5thick}{(12.5thick,1.5thick)} \end{fmfgraph*} \end{fmffile}} \;,
\end{equation}
where now the propagators are connected and the three-point vertex is the 1PI one. The matrices $\mx X$ and $\mx Y$ have been introduced and depict the expressions that can be found in the explicit equations above. These equations can be compared to the standard form of Schwinger--Dyson equations.

The matrix $\mx X$ would ordinarily be the bare Euclidean propagator $m^2+p^2$, eventually complicated with some Lorentz index structure. This is quite similar to what we have here for the charge conjugation odd sector in equation \eqref{minmin}, with the main difference that, on the lattice, momenta are replaced by sines of momenta. The charge conjugation even sector has, at this point, no momentum dependence in its bare propagator yet, but the inclusion of more complex Wilson loops beyond the elementary plaquette immediately remedies that.\footnote{Also, the momentum dependent term in the charge conjugation odd sector has the wrong sign, which also drastically changes when including the mixing with other Wilson loop shapes.} The second term in the above equation depicts mixing with other fields. In order to extract the physics in the large-$N$ limit one should diagonalize the full propagator matrix, containing all mixings of all different shapes of Wilson loops. The $\mx Y$ tensor would, in ordinary Schwinger--Dyson equations, be a Dirac delta function in position space, or a constant when going to Fourier space. In our case, it has some nontrivial momentum dependence due to the mismatch between which fields are in the path integral (the link matrices) and which fields we are considering $n$-point functions of (Wilson loops built from link matrices). The last term in the equations is a loop correction due to a three-point interaction, governed by a coupling strength $1/N$. Naively one might, at this point, be led to believe that the effective theory describing glueballs will only have a three-point vertex. Explicit consideration of higher $n$-point functions (for which see the following section) shows, however, that matters are more complicated. This can already be forseen due to the $1/N^2$ corrections the $\mx Y$ matrix receives (see the equations \eqref{tweeprops}), which superficially look like a tadpole correction from a four-point vertex, but which, all things considered, do not have the correct form to be exactly that.

\subsection{Higher $n$-point functions}
\subsubsection{Notational conventions}
In order to discuss higher $n$-point functions in a more structured way, I will introduce some matricial notation. Let us consider the Schwinger--Dyson equation for the $n$-point function containing the Wilson loops $A(0)$, $B(x)$, $C(y)$, etc. The functional derivative is taken at links of the loop $A$, which is cut upen by a gauge generator as necessary.

First let us suppose we fixed some choice determining how the functional derivatives are going to be taken. This consists of choosing the links of $A$ we will use, and a linear combination to sum the different contributions. Let us furthermore suppose that the coefficients in the linear combination add up to $-2/N$ at leading order in $1/N$,\footnote{If this is not the case, one has but to multiply the coefficients with a suitable number to adjust it. This will only affect the equations in multiplying them with a constant. Only when a certain choice leads to a zero sum of coefficients will one encouter a problem in doing so. I will in the present work ignore this degenerate case.} so chosen to have less clutter later on. Let us now consider the different terms appearing in the path integral.

First there are the different ways the derivative can work on $A(0)$. This will lead to contributions containing just $A(0)$, and contributions where the loop $A$ falls apart into two different loops, analogous to what is shown in \eqref{dubbelvaltuitelkaar}. Whichever choice one made concerning the links the functional derivative works on, the coefficient of the contribution containing just $A(0)$ will always be one at leading order for large $N$, due to the scale of the linear combination chosen.\footnote{Subleading terms of this coefficient can always be made to disappear by adjusting the linear combination.} To describe the contributions from the loop $A$ falling apart into two loops $J$ and $K$, let us introduce the notation $\alpha$ as:
\begin{equation}
\frac1N \sum_{J(j),K(k)} \alpha_{A(0),J(j),K(k)} J(j) K(k) \;.
\end{equation}
The symbol $\alpha$ may, of course, depend nontrivially on $N$, but with the definition above its leading behavior at large $N$ will be of order $N^0$. The symbol $\alpha$ can always be defined to be symmetric in its two last indices, but this is not strictly necessary.

If one of the other Wilson loops under consideration $B$, $C$, etc. is well-positioned, the functional derivatives may work on it as well and give two more kinds of terms. The first kind is where the two loops $A$ and, for definiteness say, $B$ are left as they are. These will be written as\footnote{The difference between putting the indices above or below has no specific meaning, but is chosen such that a semblance of the Einstein summation convention can be applied to indices put below, if one interprets condensates to have a lower index.}
\begin{equation}
-\frac1{N^2} A(0) B(x) \beta^{A(0),B(x)} \;,
\end{equation}
where $\beta$ is again order $N^0$ at large $N$, and where the sign is chosen to have $\beta$ be positive if the links of $A$ and $B$ hit by the functional derivative go in the same direction. The second kind are those where the loops $A$ and $B$ are linked with each other to form one larger loop $J$:
\begin{equation}
\frac1N \sum_{J(j)} \gamma_{A(0),B(x),J(j)} J(j) \;.
\end{equation}
It is possible for $A$ and $B$ to cancel each other and give the unit operator, as happens whenever they follow the same loop at the same spacetime point but have opposite orientation. This will not be included in the $\gamma$ symbol and will instead be treated separately.

Finally there will be contributions from the functional derivative working on the action, bringing down elementary plaquettes. These can be left unlinked with $A$:
\begin{equation}
-\frac1{N\lambda} A(0) \sum_{P(p)} \epsilon^{A(0)}_{P(p)} P(p) \;,
\end{equation}
or linked with $A$:
\begin{equation}
\frac1\lambda \sum_{J(j)} \zeta_{A(0),J(j)} J(j) \;.
\end{equation}

The notation $\delta_{A(x),B(y)}$ will be used as a Kronecker delta for Wilson loops, and it is one if $A$ and $B$ are identical loops taken in the same direction, and at the same spacetime point --- and zero otherwise. Furthermore propagators and vertices will from now on be written in matrix-like notation, as $\Delta_{A(x),B(y)}$ and $\Gamma_{A(x),B(y),C(z)}$.

As an example, the equation for the one-point function of the Wilson loop $A$ will now be written as
\begin{multline}
A + \sum_{J(j),K(k)} \alpha_{A(0),J(j),K(k)} J K + \frac1\lambda \sum_{J(j)} \gamma_{A(0),J(j)} J \\ + \frac1{N^2} \sum_{J(j),K(k)} \alpha_{A(0),J(j),K(k)} \Delta_{J(j),K(k)} - \frac1{N^2\lambda} \sum_{P(p)} \epsilon^{A(0)}_{P(p)} \Delta_{A(0),P(p)} = 0 \;.
\end{multline}
For the propagator equation, and also for further use, it comes in handy to define the following matrices:
\begin{subequations} \label{xeny} \begin{align}
\mx X_{A(x),B(y)} =& \delta_{A(x),B(y)} + \sum_{J(j)} \alpha_{A(x),J(j),B(y)} J + \sum_{K(k)} \alpha_{A(x),B(y),K(k)} K \nonumber \\ & - \frac1\lambda A \sum_{P(p)} \epsilon^{A(x)}_{P(p)} \delta_{P(p),B(y)} + \frac1\lambda \zeta_{A(x),B(y)} \;, \\
\mx Y_{A(x),B(y)} =& - AB \beta^{A(x),B(y)} + \sum_{J(j)} \gamma_{A(x),B(y),J(j)} J - \delta_{A(x),B^*(y)} \;,
\end{align} \end{subequations}
and the propagator equation becomes very simply:
\begin{multline}
\sum_{J(j)} \mx X_{A(0),J(j)} \Delta_{J(j),B(x)} + \mx Y_{A(0),B(x)} + \frac1N \sum_{J(j),K(k)} \alpha_{A(0),J(j),K(k)} \Gamma^\text{conn}_{J(j),K(k),B(x)} \\ - \frac1{N\lambda} \sum_{P(p)} \epsilon^{A(0)}_{P(p)} \Gamma^\text{conn}_{A(0),P(p),B(x)} - \frac1{N^2} \Delta_{A(0),B(x)} \beta^{A(0),B(x)} = 0 \;.
\end{multline}
This means that the inverse propagator at tree level (i.e. in the large-$N$ limit) is equal to $-\mx Y^{-1}\mx X$. Mark that $\mx Y$ has to be invertible for this to exist. It can easily be seen to be so in the limit of large $\lambda$, as the matrix reduces to $-\delta_{A(x),B^*(y)}$. I will in the following always suppose it to be invertible for low $\lambda$ as well, although I have no formal proof of this.

\subsubsection{Three-point functions}
The procedure to write down the equations for the connected three-point vertices is completely similar to the one for propagators. When one goes from connected to 1PI, however, two things are out of the ordinary.

As is normal in Schwinger--Dyson equations, one of the terms in the equations for the connected three-point functions is
\begin{equation}
\sum_{J(j)} \mx X_{A(0),J(j)} \Gamma^\text{conn}_{J(j),B(x),C(y)} \;.
\end{equation}
When writing the connected vertex in terms of the 1PI vertex and propagators, one gets for this term:
\begin{equation} \label{driepuntmaster}
\mx X \parbox{8.2em}{\begin{fmffile}{driepunteenpi} \begin{fmfgraph*}(8,5) \fmfleft{l} \fmfright{r1,r2} \fmf{plain,tension=2.2}{l,v3,v} \fmf{plain}{v,v1,r1} \fmf{plain}{v,v2,r2} \fmfv{decor.shape=circle,decor.filled=full,decor.size=5thick}{v1} \fmfv{decor.shape=circle,decor.filled=full,decor.size=5thick}{v2} \fmfv{decor.shape=circle,decor.filled=full,decor.size=5thick}{v3} \fmfv{decor.shape=circle,decor.filled=full,decor.size=5thick}{v} \end{fmfgraph*} \end{fmffile}} ,
\end{equation}
Then, the product $\mx X \Delta$, together with all loop corrections that are not 1PI, can be replaced by their value from the propagator equation. In the ordinary Schwinger--Dyson formalism, that would be a delta function, leaving just the 1PI vertex with two propagators sticking out. In the case at hand, a $\mx Y$ matrix appears instead.

A second divergence from the ordinary Schwinger--Dyson formalism consists in deleting the propagators sticking out to the right in \eqref{driepuntmaster}. Most terms in the three-point equation have exactly the same two propagators sticking out, as is ordinarily the case, such that those just get deleted to come to the 1PI equations. However, there are also some terms with the form
\begin{equation} \label{ambetante}
-\frac1N \quad \parbox{6.3em}{\begin{fmffile}{driepunta} \begin{fmfgraph*}(6,5) \fmfleft{l1,l2,l,l3,l4} \fmfright{r1,r2} \fmf{plain}{r2,v,r1} \fmf{plain}{l,l3} \fmf{dashes}{l,v2,r1} \fmfv{decor.shape=circle,decor.filled=full,decor.size=5thick}{v} \fmfv{decor.shape=circle,decor.filled=full,decor.size=5thick}{l3}
\fmfv{decor.shape=circle,decor.filled=empty,decor.size=7thick,label=$\beta$,label.dist=0em}{v2} \end{fmfgraph*} \end{fmffile}} - \frac1N \ \parbox{6.3em}{\begin{fmffile}{driepuntj} \begin{fmfgraph*}(6,5) \fmfleft{l} \fmfright{r1,r2} \fmf{plain}{l,v,r2} \fmf{phantom}{r1,s,r2} \fmf{plain}{r1,s} \fmf{dashes}{l,v2,r1} \fmfv{decor.shape=circle,decor.filled=full,decor.size=5thick}{v} \fmfv{decor.shape=circle,decor.filled=full,decor.size=5thick}{s}
\fmfv{decor.shape=circle,decor.filled=empty,decor.size=7thick,label=$\beta$,label.dist=0em}{v2} \end{fmfgraph*} \end{fmffile}} + \frac 1N \ \parbox{6em}{\begin{fmffile}{driepuntgamma} \begin{fmfgraph*}(6,5) \fmfleft{l} \fmfright{r1,r2} \fmf{dashes}{l,v,r1} \fmffreeze \fmf{plain}{v,v2,r2} \fmfv{decor.shape=circle,decor.filled=empty,decor.size=7thick,label=$\gamma$,label.dist=0em}{v} \fmfv{decor.shape=circle,decor.filled=full,decor.size=5thick}{v2} \end{fmfgraph*} \end{fmffile}}
\end{equation}
plus the same with the two rightmost legs interchanged. The circles with $\beta$ and $\gamma$ inside depict the symbols defined above. These terms do not have the two propagators sticking out at the right, meaning that they will eventually get inverse propagators multiplying them. If we are interested in the large-$N$ values, this inverse propagator can be replaced by its tree-level value of $-\mx Y^{-1} \mx X$.

The final result in the large-$N$ limit is
\begin{multline} \label{driepuntvgl}
\sum_{J(j)} \mx Y_{A(0),J(j)} \Gamma^\text{1PI}_{J(j),B(x),C(y)} = \frac1N \alpha_{A(0),B(x),C(y)} - \frac1N A \beta^{A(0),C(y)} \Delta^{-1}_{B(x),C(y)} \\ - \frac1N \delta_{A(0),C(y)} \sum_{J(j)} J \beta^{A(0),J(j)} \Delta^{-1}_{J(j),B(x)}  + \frac1N \sum_{J(j)} \gamma_{A(0),J(j),C(y)} \Delta^{-1}_{J(j),B(x)} \\ - \frac1{N\lambda} \delta_{A(0),B(x)} \sum_{P(p)} \epsilon^{A(0)}_{P(p)} \delta_{P(p),C(y)} + (B(x)\leftrightarrow C(y)) + \mathcal O(\tfrac1{N^3}) \;.
\end{multline}
This also confirms that the three-point vertex has order $1/N$.

\subsubsection{Four-point functions and higher}
The issues encountered when writing down the equations for higher $n$-point functions are similar to the ones of the three-point vertex. When going from the equations for the connected vertices to the ones for the 1PI vertices, one has to use the equations for the three-point vertex to get rid of certain terms that are not 1PI, but the terms like the ones in \eqref{ambetante} do not match and introduce some extra terms. These look like
\begin{equation}
\frac1N \quad \parbox{8.3em}{\begin{fmffile}{vierpunta} \begin{fmfgraph*}(8,6) \fmfleft{l1,l2,l,l3,l4} \fmfright{r1,r2,r3} \fmf{plain}{r3,b1,v1} \fmf{plain}{r2,b2,v2} \fmf{plain}{r1,b3,v2} \fmf{plain,tension=3}{v1,v2} \fmf{plain}{l,l3} \fmf{dashes,tension=4}{l,v3,v1} \fmfv{decor.shape=circle,decor.filled=full,decor.size=5thick}{l3} \fmfv{decor.shape=circle,decor.filled=full,decor.size=5thick}{b1} \fmfv{decor.shape=circle,decor.filled=full,decor.size=5thick}{b2}
\fmfv{decor.shape=circle,decor.filled=full,decor.size=5thick}{b3} \fmfv{decor.shape=circle,decor.filled=full,decor.size=5thick}{v2} \fmfv{decor.shape=circle,decor.filled=empty,decor.size=7thick,label=$\beta$,label.dist=0em}{v3} \end{fmfgraph*} \end{fmffile}} + \frac1N \ \parbox{8.3em}{\begin{fmffile}{vierpuntj} \begin{fmfgraph*}(8,6) \fmfleft{l1,l,l2} \fmfright{r1,r2,r3} \fmf{plain}{l,b1,r3} \fmf{plain,tension=3}{v1,v2} \fmf{plain}{v2,b2,r2} \fmf{plain}{v2,b3,r1} \fmf{dashes,tension=2}{l,be,v1} \fmffreeze \fmf{phantom}{v1,s,l1} \fmf{phantom,tension=.5}{s,r1} \fmf{plain,tension=0}{v1,s} \fmfv{decor.shape=circle,decor.filled=full,decor.size=5thick}{b1} \fmfv{decor.shape=circle,decor.filled=full,decor.size=5thick}{b2} \fmfv{decor.shape=circle,decor.filled=full,decor.size=5thick}{b3} \fmfv{decor.shape=circle,decor.filled=full,decor.size=5thick}{v2} \fmfv{decor.shape=circle,decor.filled=full,decor.size=5thick}{s} \fmfv{decor.shape=circle,decor.filled=empty,decor.size=7thick,label=$\beta$,label.dist=0em}{be} \end{fmfgraph*} \end{fmffile}} - \frac1N \ \parbox{8em}{\begin{fmffile}{vierpuntgamma} \begin{fmfgraph*}(8,6) \fmfleft{l} \fmfright{r1,r2,r3} \fmf{dashes,tension=1.5}{l,g,v1} \fmf{plain}{v1,b2,r2} \fmf{plain}{v1,b3,r1} \fmffreeze \fmf{plain}{g,b1,r3} \fmfv{decor.shape=circle,decor.filled=empty,decor.size=7thick,label=$\gamma$,label.dist=0em}{g} \fmfv{decor.shape=circle,decor.filled=full,decor.size=5thick}{v1} \fmfv{decor.shape=circle,decor.filled=full,decor.size=5thick}{b1} \fmfv{decor.shape=circle,decor.filled=full,decor.size=5thick}{b2} \fmfv{decor.shape=circle,decor.filled=full,decor.size=5thick}{b3} \end{fmfgraph*} \end{fmffile}}
\end{equation}
and their permutations. Due to these terms, the tree-level value of the four-point vertex will depend on the three-point vertex. This will carry over to higher $n$-point vertices. The final result for the four-point vertex for large $N$ is
\begin{multline}
\sum_{J(j)} \mx Y_{A(0),J(j)} \Gamma^\text{1PI}_{J(j),B(x),C(y),D(z)} = -\frac1{N^2} \delta_{A(0),D(z)} \beta^{A(0),C(y)} \Delta^{-1}_{B(x),C(y)} \\ + \frac1N A \beta^{A(0),B(z)} \Gamma^\text{1PI}_{B(z),C(y),D(z)} + \frac1N \delta_{A(0),B(x)} \sum_{J(j)} J \beta^{A(0),J(j)} \Gamma^\text{1PI}_{J(j),C(y),D(z)} \\ - \frac1N \sum_{J(j)} \gamma_{A(0),J(j),B(x)} \Gamma^\text{1PI}_{J(j),C(y),D(z)} + (\text{permutations of $B(x)$, $C(y)$ and $D(z)$}) + \mathcal O (\tfrac1{N^4}) \;.
\end{multline}
The four-point vertex has order $1/N^2$. Higher $n$-point vertices will have order $1/N^{n-2}$.

\section{Towards concrete results} \label{tcr}
In this section I present several concrete results one can get out of the equations formulated earlier. A full solutions is, even in the large-$N$ limit, not straightforward to obtain. For example, in order to obtain the spectrum, one should compute the matrices $\mx X$ and $\mx Y$ in \eqref{xeny}, after which the product $-\mx Y^{-1}\mx X$ (or its inverse) should be diagonalized. As both matrices are infinitely big, this is a rather formidable task to do with only brute force. And without having at least the inverse of the propagator, interaction vertices cannot even be considered.

One regime that is readily accessible with only brute force is the limit of large 't~Hooft coupling $\lambda$. The physical limit is, however, the one of $\lambda$ going to zero, as this will be the continuum limit.\footnote{As I have not spoken a word of renormalization, $\lambda$ is always the bare coupling, which is linked to the lattice spacing through the renormalization group equations.} In the large-$N$ limit, there appears to be a phase transition when going from large to small $\lambda$, making a strong coupling expansion less usefull as a way to approach the physical limit \cite{Gross:1980he}. Furthermore, one has to go to rather high order to see any difference between different numbers of spacetime dimensions, which means that the rather trivial results of two-dimensional pure Yang--Mills theory will be found until one can gather the patience to go through the ever more complex higher-order computations.

Another case that is partially accessible is the one of two spacetime dimensions. Again it is not physically the most useful case, but it can serve as a playground to test the formalism.

A possible way to simplify the task of extracting physics is by doing a change of variables in field space. It seems to be possible to get rid of all inverses of $\mx Y$ in the tree-level expressions for the inverse propagator and for the three-point vertex, and furthermore all the matrix products that are left become finite sums. This means that arbitrary terms in the Lagrangian density describing the effective theory can be written down in finite time without the need for further artifice. This does not yet yield the physics we are most interested in, as the propagator will still be an infinitely big matrix function of the unknown one-point functions, but it is a promising avenue for further exploration. This is discussed in paragraph \ref{boaet}

\subsection{Pure Yang--Mills in $2d$}
Let us first consider the case of two spacetime dimensions. Following the usual steps (see, for example, \cite{Gross:1980he}) it is possible to reduce the theory to contain only one link matrix:
\begin{equation}
\mathcal Z = \int [dU] e^{\frac N\lambda (\tr U+\tr U^\dagger)} \;.
\end{equation}
The observables of the theory are described by traced powers of this one link matrix:
\begin{equation} \label{defwilsonloops}
\tr U^n \;.
\end{equation}
The equations for the vacuum expectation values hereof in the large-$N$ limit have already been considered and solved in \cite{Paffuti:1980cs,Friedan:1980tu}:\footnote{In this expression, the case of $n=1$ in the $\lambda<2$ region can be obtained by taking the limit $n\to1$. The explicit result is $W_1 = 1-\tfrac14\lambda$.}
\begin{equation} \label{wilsontweed}
W_n = \langle \tfrac1N \tr U^n \rangle = \begin{cases} (1-\tfrac12\lambda) \left(\frac{P'_n(1-\lambda)}{n(n+1)} + \frac{P'_{n-1}(1-\lambda)}{n(n-1)}\right) & \quad (\lambda < 2) \\ \frac1\lambda \delta_{n,1} & \quad (\lambda > 2) \end{cases} \;,
\end{equation}
where the $P_n(x)$ are the Legendre polynomials. At the point $\lambda=2$ a third-order phase transition happens \cite{Gross:1980he}. The $W_n$ are once continuously differentiable at that point.

For $\lambda>2$, the system of equations for the propagators simplifies considerably due to most one-point functions being zero. After splitting the system according to charge conjugation, the solution can be readily found to be diagonal in both sectors:
\begin{equation}
\Delta^{++}_{mn} = \frac m2 \delta_{m,n} \;, \qquad \Delta^{--}_{m,n} = - \frac m2 \delta_{mn} \;.
\end{equation}
Interaction vertices all turn out to be zero. Mark, however, that the absence of interactions at tree level does not imply the absence of $1/N^2$ corrections, as the bare propagator, and indeed even the vertices, receive ``finite renormalizations'', thus leading to non-trivial behavior at nonleading order. Indeed, explicitely going through the equations to compute the corrections uncovers, for example, a nonzero correction to the one-point functions. In order to compute this correction, one can use a strong-coupling series as an ansatz for the correction:
\begin{equation}
\frac1{N^2} \frac1{\lambda^n} \sum_{i=0}^\infty \alpha_{n,i} \frac1{\lambda^{2i}} \;,
\end{equation}
after which a recursion formula is obtained for the coefficients $\alpha_{n,i}$. This recursion formula can be recognized to be the one for the Catalan triangle in slightly shifted form. The one-point functions are then:
\begin{equation}
W_n = \frac1\lambda \delta_{n,1} + \frac1{N^2} \left(\frac1\lambda \delta_{n,1} + \left(\sqrt{\frac{\lambda^2}4 - 1} - \frac\lambda2\right)^n\right) + \mathcal O(\tfrac1{N^4}) \;.
\end{equation}

The case of $\lambda<2$ is more complex, and it is not possible to solve the equations merely by squinting at them. Instead one can use the technique of the generating function \cite{Paffuti:1980cs,Friedan:1980tu} to rewrite the equations in term of the unknown generating function of the propagators $\sum_{m,n} x^m y^n \Delta_{mn}$. In order to solve this algebraic equation, one needs the input $\Delta_{1n}$. In the charge-conjugation even sector one happens to have the identity
\begin{equation}
\Delta_{1n}^{++} = -\frac{\lambda^2}2 \frac d{d\lambda} W_n \;,
\end{equation}
which follows from deriving the path integral with respect to $\lambda$. The result can be found to be
\begin{equation}
\Delta^{++} = \begin{pmatrix} \frac{\lambda^2}8 & \frac{\lambda^2}2 - \frac{\lambda^3}4 & \frac{9 \lambda^2}8 - \frac{3\lambda^3}2 + \frac{15\lambda^4}{32} & \cdots \\ \frac{\lambda^2}2 - \frac{\lambda^3}4 & 2\lambda^2 - 2 \lambda^3 + \frac{9\lambda^4}{16} & \frac{9\lambda^2}2 - \frac{33\lambda^3}4 + \frac{21\lambda^4}4 - \frac{9\lambda^5}8 & \cdots \\ \frac{9 \lambda^2}8 - \frac{3\lambda^3}2 + \frac{15\lambda^4}{32} & \frac{9\lambda^2}2 - \frac{33\lambda^3}4 + \frac{21\lambda^4}4 - \frac{9\lambda^5}8 & \frac{81\lambda^2}8 - 27 \lambda^3 + \frac{459\lambda^4}{16} - \frac{27\lambda^5}2 + \frac{75\lambda^6}{32} & \cdots \\ \vdots & \vdots & \vdots & \ddots \end{pmatrix} .
\end{equation}
This connects continuously (but with discontinuous derivative) to the result for $\lambda>2$. I have not yet found any way to compute the charge-conjugation odd propagator.

The charge-conjugation even connected vertices can be computed in the same way. The three-vertex turns out to factorize:
\begin{equation}
\Gamma^{+++}_{lmn} = \frac1N \gamma_l\gamma_m\gamma_n \;, \qquad \gamma_n = -\frac\lambda2 (P_n'(1-\lambda)+P_{n-1}'(1-\lambda)) \;.
\end{equation}
It is discontinuous at $\lambda=2$. Extracting the 1PI vertex would involve inverting the propagator, or alternatively (if one starts from the equation \eqref{driepuntvgl} for the 1PI vertex) inverting the $\mx Y$ matrix, neither which I have been able to do so far.

This rather trivial example of two-dimensional pure Yang--Mills theory shows the soundness of the formalism. From the equations, the symmetry of the $n$-point functions under interchange of their legs is not at all trivial, as one leg is alway specially singled out. The results cleanly recover the necessary symmetry. The example also shows the difficulty in extracting physical information, as infinitely big matrices have to be inverted. In the higher-dimensional case, this will only be worse, as it will not be possible anymore to reduce the theory to as simple a set of fields as in \eqref{defwilsonloops}.

\subsection{Beginnings of an effective theory} \label{boaet}
It turns out a certain simplification happens when performing the following change of variables in the effective action:
\begin{equation} \label{verandervar}
A(x) = \sum_{B'(y)} \mx Y_{B'(y),A(x)} B'(y) \;,
\end{equation}
where the fields with a prime are the new ones. For the 1PI $n$-point functions this results in the change
\begin{subequations} \begin{gather}
\Delta'^{-1}_{A(x),B(y)} = \sum_{J(j),K(k)} \mx Y_{A(x),J(j),} \mx Y_{B(y),K(k)} \Delta^{-1}_{J(j),K(k)} \;, \\
\Gamma'_{A(x),B(y),C(z)} = \sum_{J(j),K(k),L(l)} \mx Y_{A(x),J(j)} \mx Y_{B(y),K(k)} \mx Y_{C(z),L(l)} \Gamma_{J(j),K(k),L(l)} \;,
\end{gather} \end{subequations}
and so on for higher vertices. The new inverse propagator then becomes at leading order in $1/N$:
\begin{equation}
\Delta'^{-1}_{A(0),B(x)} = - \sum_{J(j)} \mx X_{A(0),J(j)} \mx Y_{B(x),J(j)} + \mathcal O(\tfrac1{N^2}) \;.
\end{equation}
Not only is it no longer necessary to invert an infinitely big matrix, but furthermore the sum over Wilson loops $J$ is finite, as $\mx X_{A(0),J(j)}$ is, for any $A$, only nonzero for a limited number of $J(j)$. With the above formula, one can, in principle, write down any propagator term from the effective Lagrangian describing glueball dynamics.

One could argue that the change in variables \eqref{verandervar} is, in general, highly nonlocal, thus possibly introducing spurious zeros into the inverse propagator, or a nonphysical particle into the theory. This indeed happens, as can already be verified by considering the two-dimensional theory without reducing it to one link (i.e. two-dimensional pure Yang--Mills theory in its most naive form). In the charge-conjugation even sector, the inverse propagator for the operator $P'_{12}(x)$ turns out to be
\begin{equation}
(\Delta'^{-1})^{++}_{P',P'}(p) = 2(1+3W_1^2+W_2) - 2W_1^2 \sum_\mu \sin^2\tfrac{p_\mu}2 \;.
\end{equation}
If this is compared with the inverse propagator of a free, massive scalar particle on the lattice
\begin{equation}
m^2 + \sum_\mu \sin^2\tfrac{p_\mu}2 \;,
\end{equation}
it looks like a physical particle must be present in the theory,\footnote{Including mixing terms with other fields may lead to a more complex picture, but it will not make the pole in the propagator disappear.} while we know that the two-dimensional theory with only glue has an empty spectrum. In order to be able to discern between real particles and fictitious particles introduced by the change of variables \eqref{verandervar}, one has to take into account the interactions between the particles, as such a fictitious particle will have no on-shell interactions.

With the above change in variables, it is also possible to compute elements of the three-point vertex at tree level:
\begin{multline}
\Gamma'_{A(0),B(x),C(y)} = \frac1N \sum_{J(j),K(k)} \alpha_{A(0),J(j),K(k)} \mx Y_{B(x),J(j)} \mx Y_{C(y),K(k)} \\ + \frac1N A \sum_{J(j)} \beta^{A(0),J(j)} \mx X_{B(x),J(j)} \mx Y_{C(y),J(j)} + \frac1N \sum_{J(j)} J \beta^{A(0),J(j)} \mx X_{B(x),J(j)} \mx Y_{C(y),A(0)} \\ - \frac1N \sum_{J(j),K(k)} \gamma_{A(0),J(j),K(k)} \mx X_{B(x),J(j)} \mx Y_{C(y),K(k)} - \frac1{N\lambda} \sum_{P(p)} \epsilon^{A(0)}_{P(p)} \mx Y_{B(x),A(0)} \mx Y_{C(y),P(p)} \\ + (B(x) \leftrightarrow C(y)) + \mathcal O(\tfrac1{N^3}) \;.
\end{multline}
Again, whatever be $A$, all sums will be finite. This allows one to write down any cubic term in the effective Lagrangian with a finite number of computations. If one applies this to the case of the $P'_{12}(x)$ operator in two dimensions, one finds that the vertex is also momentum dependent, thus leaving open the possiblity of the vertex to have a zero whenever the inverse propagator has a zero. In order to verify this, however, the full inverse propagator should be known, thus still calling for more powerful techniques to tackle the problem.

The formula for higher $n$-point functions does not undergo the same simplification.

\section{Conclusions and outlook} \label{outlook}
In this paper, Schwinger--Dyson equations for correlators of Wilson loops were developed. It turns out to be possible to write down equations for glueball propagators and interactions, and the coupling strength governing loop corrections is the inverse number of colors $1/N$. This means that, in the large-$N$ limit, the equations simplify to equations for tree-level $n$-point functions, and $1/N$ corrections can be systematically written down.

In order to extract physics, however, more work will still be needed. Even at tree level the tower of equations is infinitely large, and only in the trivial case of two-dimensional pure Yang--Mills can the solution be partially extracted without too many additional difficulty. In all other cases, infinitely big and rather unwieldy matrices have to be handled in order to find the results that interest us. These matrices furthermore depend on the one-point functions, which have to be solved for --- a task that has been successfully completed only in a very limited number of settings \cite{Friedan:1980tu}, none of them resembling the real world.

On the other hand, the formalism has some as yet unexplored strength. Before writing down any equation, one must choose at what sites to take the functional derivative. This has no equivalent in ordinary Schwinger--Dyson equations and is due to the field variables in the path integral differing from the one we compute correlation functions of. Due to this choice, one has the liberty to search for different recastings of the equations, and it may very well be that a solution would come within reach if only the right choice is made. This will be left for further research.

Another important avenue of further research would be the inclusion of quarks. Due to their fermionic nature, however, it is less natural to treat quarks on the lattice --- due to the fermion doubling problem. Several techniques exist to avoid this problem, but always at the cost of increased complexity.

\section*{Acknowledgements}
This work is supported by the Generalitat Valenciana under grant Prometeo/2008/004 and by the Spanish MICINN under grant FPA2011-23596.

\bibliographystyle{unsrt}
\bibliography{Bibliografie}

\end{document}